\documentclass[12pt,twoside]{article}
\usepackage[dvips]{color}
\usepackage{textcomp}
\usepackage{amsmath}
\usepackage{amssymb}
\usepackage{graphicx}
\usepackage{floatflt}
\usepackage{wrapfig}
\usepackage[colorlinks=true]{hyperref}

\newlength{\figwid}
\newcommand{\aerr}[2]{\ensuremath{\mathrel{\hbox{\rlap{\hbox{\lower2.4pt\hbox{\scriptsize{#2}}}}\rai
se4.3pt\hbox{\scriptsize{#1}}}}}}
\newcommand{\aberr}[2]{\ensuremath{\mathrel{\hbox{\rlap{\hbox{\lower4.4pt\hbox{\small{#2}}}}\raise4.
4pt\hbox{\small{#1}}}}}}

\def\issue(#1,#2,#3){#1 (#3) #2} 

\def\opcit(#1){ {\em op. cit.}, #1}
\def\NIM(#1,#2,#3){ Nucl.\ Instrum.\ and Meth.\ \issue(#1,#2,#3)}

\definecolor{pink}{cmyk}{.5,.6,.2,.1}
\definecolor{dgreen}{cmyk}{.7,.1,.7,.2}
\definecolor{orange}{cmyk}{.0,.5,1.,.0}
\definecolor{brown}{rgb}{0.7,0.3,0.0}
\definecolor{purple}{rgb}{0.6,0.1,0.5}

\begin{document}

\begin{center}
\large
A more exact solution for incorporating multiplicative systematic uncertainties in branching ratio limits
\end{center}

\begin{flushleft}
Kevin Stenson,
Department of Physics,
University of Colorado,
UCB 390,
Boulder, CO 80309\\
\texttt{stenson@fnal.gov}
\end{flushleft}

A method for incorporating systematic errors into branching ratio limits which are not obtained from
a simple counting analysis has been suggested by Mark Convery~\cite{convery}.  The derivation makes 
some approximations which are not necessarily valid.  This note presents the full solution as an
alternative.  The basic idea is a simple
extension of the Cousins and Highland philosophy~\cite{cousins}.  Before systematics are considered,
an analysis using a maximum likelihood fit returns a central value for the branching ratio $(\hat{B})$
and a statistical error $(\sigma_B)$.  The likelihood function is
\begin{equation}
p(B) \propto \exp{\!\left[\frac{-(B-\hat{B})^2}{2 \sigma_B^2}\right]}.
\end{equation}
Following the Convery notation, we associate $\hat{S}$ with the nominal efficiency and $\sigma_S$ as
the (Gaussian) error on the efficiency.  Adding the uncertainty on the efficiency changes the likelihood to:
\begin{equation}
\label{eq:integral}
p(B) \propto \int_0^1 \exp{\!\left[\frac{-(S B/\hat{S} - \hat{B})^2}{2 \sigma_B^2}\right]} 
\exp{\!\left[\frac{-(S-\hat{S})^2}{2\sigma_s^2}\right]} d\!S.
\end{equation}
From Mathematica$^\textrm{\textregistered}$, the integral in Eq.~\ref{eq:integral} is:
\small
\begin{equation*}
\label{eq:full}
\sqrt{\frac{\pi}{2}}\frac{\hat{S}}{\sqrt{\frac{B^2}{\sigma_B^2}+\frac{\hat{S}^2}{\sigma_S^2}}}
\exp{\!\!\left[\!\frac{-(B-\hat{B})^2}{2\left(\frac{B^2\sigma_S^2}{\hat{S}^2}+\sigma_B^2\right)}\!\right]}\!
\left\{\textrm{erf}\!\left[\!\frac{\hat{S}\left(\sigma_B^2 + \frac{B \hat{B}\sigma_S^2}{\hat{S}^2}\right)}{\sqrt{2}\sigma_B\sigma_S\sqrt{\frac{B^2\sigma_S^2}{\hat{S}^2}+\sigma_B^2}}\!\right] -
\textrm{erf}\!\left[\!\frac{(\hat{S}-1)\sigma_B^2 - B \sigma_S^2\left(\frac{b}{\hat{S}^2}-\frac{\hat{B}}{\hat{S}}\right)}{\sqrt{2}\sigma_B\sigma_S\sqrt{\frac{B^2\sigma_S^2}{\hat{S}^2}+\sigma_B^2}}\!\right]\!\right\}\!.
\end{equation*}
\normalsize
Removing unimportant multiplicative constants
and changing variables from $\sigma_S$ to $\sigma_\epsilon \equiv \sigma_S/\hat{S}$ gives:
\small
\begin{equation}
\label{eq:fulleps}
p(B) \propto \frac{1}{\sqrt{\frac{B^2}{\sigma_B^2}+\frac{1}{\sigma_\epsilon^2}}}
\exp{\!\!\left[\!\frac{-(B-\hat{B})^2}{2(B^2\sigma_\epsilon^2+\sigma_B^2)}\!\right]}\!\left\{
\textrm{erf}\!\left[\!\frac{B \hat{B} \sigma_\epsilon^2 + \sigma_B^2}{\sqrt{2}\sigma_\epsilon \sigma_B \sqrt{B^2\sigma_\epsilon^2 + \sigma_B^2}}\!\right] - 
\textrm{erf}\!\left[\!\frac{(\hat{S}-1)\sigma_B^2 - B\sigma_\epsilon^2 (B - \hat{B}\hat{S})}{\sqrt{2}\hat{S}\sigma_\epsilon \sigma_B \sqrt{B^2 \sigma_\epsilon^2 + \sigma_B^2}}\!\right]\!\right\}\!.
\end{equation}
\normalsize
It turns out that as long as the efficiency $\hat{S}$ is sufficiently small (generally less than 10\% but
dependent on other parameters), the second \textbf{erf} term evaluates to $-1$ and the dependence on the
efficiency is removed.

The solution to the integral presented by Convery (for $\sigma_S \ll \hat{S}$) can be written as:
\begin{equation}
\label{eq:simp}
p(B) \propto \frac{1}{\sqrt{\frac{B^2}{\sigma_B^2}+\frac{1}{\sigma_\epsilon^2}}} 
\exp{\!\left[\frac{-(B-\hat{B})^2}{2 \left(B^2\sigma_\epsilon^2 + \sigma_B^2\right)}\right]}.
\end{equation}

The differences between Eq.~\ref{eq:fulleps} and Eq.~\ref{eq:simp} are the two 
\textbf{erf} terms in Eq.~\ref{eq:fulleps}.  The first \textbf{erf} term affects the tails of the
distribution and becomes increasingly important as $\sigma_\epsilon$ increases.  The second 
\textbf{erf} term affects the peak position and is important when $\hat{S}\pm\sigma_S$ is not easily
contained in the region $\{0,1\}$.  Or, for a fixed $\sigma_\epsilon$, when $\hat{S}$ approaches unity.
Next we compare the two results after modifying Equations~\ref{eq:simp} and \ref{eq:fulleps} to 
normalize them such that $p(B=\hat{B})=1$.

\begin{figure}[tbhp]
\includegraphics[width=\figwid]{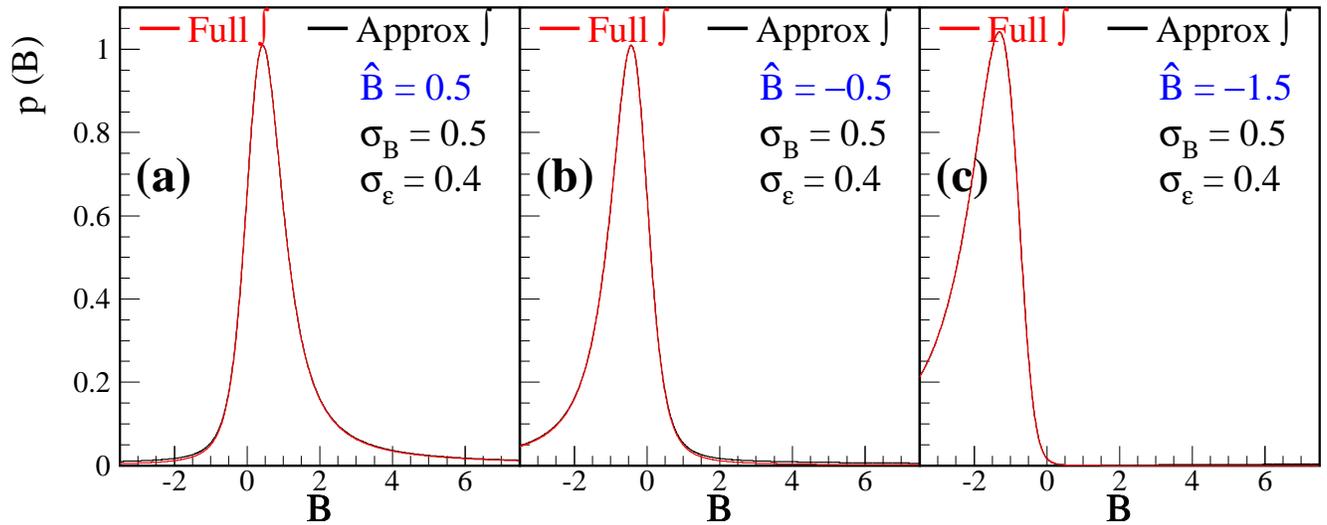}
\caption{Each plot shows a comparison of the approximate solution given by Eq.~\ref{eq:simp} in black to 
the full solution given by Eq.~\ref{eq:fulleps} in red.  For all plots, $\sigma_B=0.5$, $\sigma_\epsilon = 0.4$,
and $\hat{S}=\epsilon=0.1$.  The three plots show results for $\hat{B}=0.5$, $\hat{B}=-0.5$, and $\hat{B}=-1.5$.}
\label{fig:prob1}
\end{figure}

\begin{figure}[tbhp]
\includegraphics[width=\figwid]{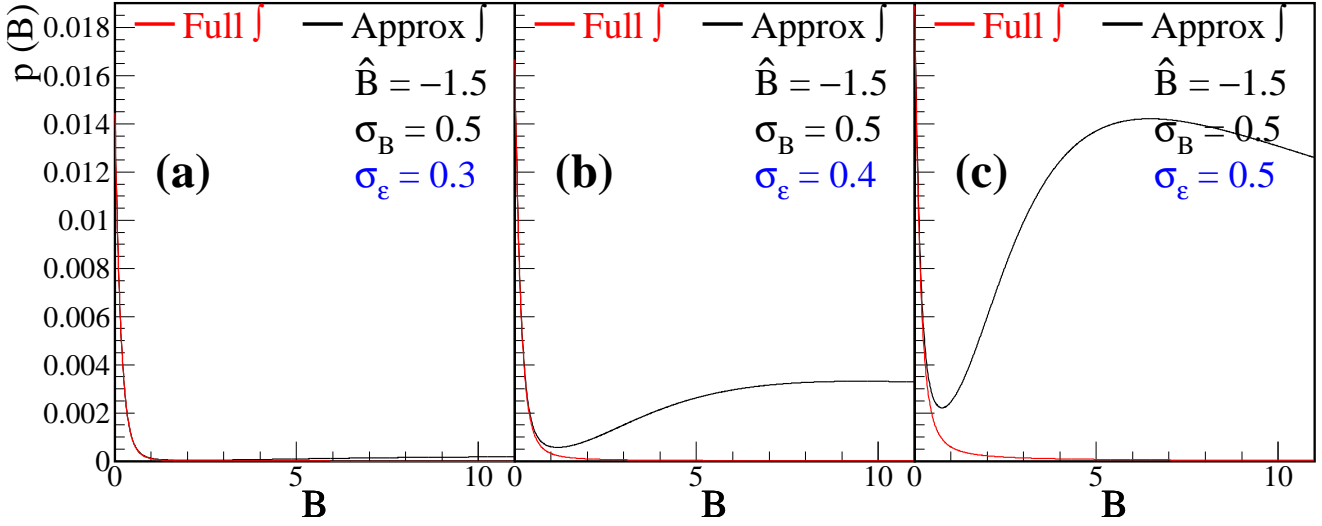}
\caption{Each plot shows a comparison of the approximate solution given by Eq.~\ref{eq:simp} in black to 
the full solution given by Eq.~\ref{eq:fulleps} in red.  For all plots, $\sigma_B=0.5$, $\hat{B} = -1.5$,
and $\hat{S}=\epsilon=0.1$.  The three plots show results for $\sigma_\epsilon=0.3$, $\sigma_\epsilon=0.4$, 
and $\sigma_\epsilon=0.5$.  In this case, the full solution is indistinguishable from the full solution
with the second \textbf{erf} term replaced by $-1$.}
\label{fig:prob2}
\end{figure}

First we check the effect for relatively large $\sigma_\epsilon$ and small $\hat{S}$ for which the 
first \textbf{erf} term becomes important.
Each plot of Figure~\ref{fig:prob1} shows a comparison between the full solution in red and the approximate
solution in black.  There is very little discernible difference between the two solutions.  The different
plots show results for $\hat{B}=0.5$, $\hat{B}=-0.5$, and $\hat{B}=-1.5$.  To set an upper limit, one
often integrates the probability over the physical region only $(B>0)$.  Figure~\ref{fig:prob2} shows the
results for $p(B)$ over the range $B\in\{0,17\}$ for the case of $\hat{B}=-1.5$ and $\sigma_B=0.5$ which 
corresponds to a $3\sigma$ negative fluctuation.  In this case clear differences between the full solution
(in red) and the approximate solution (black) can be seen for $\sigma_\epsilon \geq 0.3$.  Note that 
Fig.~\ref{fig:prob1}(a) and Fig.~\ref{fig:prob2}(b) show the same curves, only the range has changed.
Clearly an attempt to find an upper limit by integrating the area under the approximate solution is 
problematic for all the cases shown in Fig.~\ref{fig:prob2}.  Conversely, the full solution
finds an acceptable upper limit.

\begin{figure}[tbhp]
\includegraphics[width=\figwid]{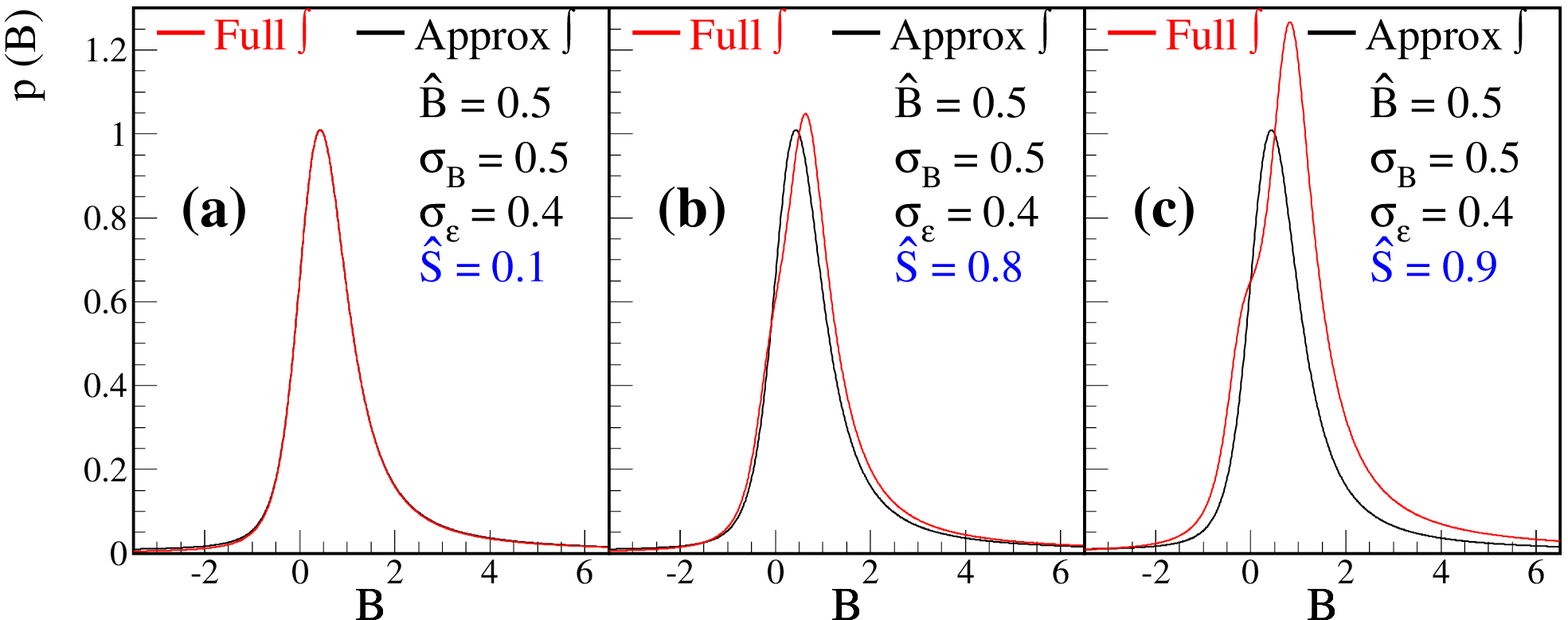}
\caption{Each plot shows a comparison of the approximate solution given by Eq.~\ref{eq:simp} in black to 
the full solution given by Eq.~\ref{eq:fulleps} in red.  For all plots, $\sigma_B=0.5$, $\hat{B} = 0.5$,
and $\sigma_\epsilon=0.4$.  The three plots show results for $\hat{S}=\epsilon=0.1$, $\hat{S}=\epsilon=0.8$,
and $\hat{S}=\epsilon=0.9$.  In this case, the full solution is nearly indistinguishable from the full solution
with the first \textbf{erf} term replaced by $+1$.}
\label{fig:prob3}
\end{figure}

Second we check the effect of the second \textbf{erf} term of Eq.~\ref{eq:fulleps} which is important 
when the integration of efficiency from $0$ to $1$ in Eq.~\ref{eq:integral} cuts off a 
significant part of the Gaussian defined by $\hat{S}\pm\sigma_S = \hat{S}\pm\sigma_\epsilon\hat{S}$.  
Figure~\ref{fig:prob3}(a) is a repeat of Fig.~\ref{fig:prob1}(a) on a different scale and again shows
little difference between the two methods.  Figures~\ref{fig:prob3}(b) and \ref{fig:prob3}(c) show the
effect of the second \textbf{erf} term as $\hat{S}\rightarrow 1$.

In conclusion, Eq.~\ref{eq:fulleps} provides a more exact and robust implementation of the original suggestion
by Convery~\cite{convery} on incorporating multiplicative systematic uncertainties in branching ratio limits.

\vspace{-4pt}

\bibliographystyle{unsrt}

\end{document}